\documentclass[english]{article}

\usepackage[T1]{fontenc}
\usepackage[latin9]{inputenc}
\usepackage{babel}
\usepackage{verbatim}
\usepackage{varioref}
\usepackage{multirow}
\usepackage{listings}

\makeatletter

\providecommand{\tabularnewline}{\\}
\newenvironment{lyxcode}
{\par\begin{list}{}{
			\setlength{\rightmargin}{\leftmargin}
			\setlength{\listparindent}{0pt}% needed for AMS classes
			\raggedright
			\setlength{\itemsep}{0pt}
			\setlength{\parsep}{0pt}
			\normalfont\ttfamily}%
		\item[]}
{\end{list}}

\newcommand{\code}[1]{\texttt{#1}}

\makeatother

\begin{document}

\title{Object-Oriented Theorem Proving (OOTP):\\
First Thoughts}

\author{Moez A. AbdelGawad\\
\medskip{}Informatics Research Institute, SRTA-City, Alexandria, Egypt\\
\code{moez@cs.rice.edu}}
\maketitle
\begin{abstract}
Automatic (i.e., computer-assisted) theorem proving (ATP) can come in
many flavors. This document presents early steps in our effort towards defining object-oriented theorem proving (OOTP) as a new
style of ATP.

Traditional theorem proving (TTP) is the only well-known flavor of ATP so far. OOTP is a generalization of TTP. While TTP is strongly based on functional programming (FP), OOTP is strongly based on object-oriented programming (OOP) instead. We believe OOTP is a style of theorem proving that is no less powerful and no less natural than TTP and thus likely will be no less practically useful than TTP.

In the document we also discuss, very briefly, a related notion of OO software verification (OOSV) based on OOTP. To clarify the relation between OOTP and TTP, we also touch on the relation between OOP and FP.
\end{abstract}

\section{Introduction}
One of the main advantages of functional programming (FP)
languages is that proving properties of most functional programs---programs written using
functional programming languages such as ML---is often straightforward.
It is well-known that traditional (automatic) theorem
proving (which we abbreviate TTP) and functional programming go hand in
hand~\cite{Paulson1996}. In fact the development of FP shows that TTP was the
original \emph{raison d'être} for FP.

To see how and why that is the case, it should be noted that TTP is typically based on tactics and tacticals. Tactics and tacticals are \emph{higher-order functions}. Tacticals are even higher-order
than tactics, since tacticals take other tactics---higher-order functions---as
their input and produce them as results.

Further, the formal verification
of the correctness of functional software is usually done using functional
software tools, so as to have a degree of mathematical confidence in the
capabilities of these proving tools (also called \emph{proof assistants}). That is, not only is the proof of the correctness of a function
in a functional program itself a function, but most of the professional
software tools used in TTP (e.g., Coq and Isabelle) are themselves---or
 at least their most important core (i.e., their ``proving engine'',
based on tactics and tacticals)---written mostly in functional languages.\footnote{That is, a functional program is used to reason about
another functional program to produce a functional program---a proof of
the correctness of the input program---as its result: we have functional
processing of functional input to produce functional output.} Hence, the
very strong connection between TTP and FP.

In this work we suggest a new style of theorem proving, object-oriented
theorem proving (OOTP), that goes hand in hand with object-oriented
programming (OOP) in the same way as TTP goes with FP. A main aim of OOTP is to make proving properties of a
large subset of object-oriented programs as equally straightforward as TTP makes it for a large subset of functional software.\footnote{This subset includes referentially-transparent, provably-terminating OO programs. As such, our work justifies OO developers adopting the so-called ``almost-functional''
style of OO programming, where developers largely avoid imperative
(i.e., non-referentially-transparent) features of OO languages (such
as mutable fields and variables).

It should be noted though that in fact this ``almost-functional''
style of OOP seems to be not necessary! As Paulson notes in~\cite[p.58]{Paulson1996},
any procedural code (including one containing \code{goto} statements and
assignment statements, the least opaque code) can be translated
%, {[}in a straightforward manner{]},
into a set of \emph{mutually recursive}
functions of a functional program. In other words, as we explain
in this work, the opaque code can be translated into an object.
(See Appendix~\ref{app:Imp-OO} for the translation of the code in~\cite[p.58]{Paulson1996} to OO code.)
Employing OOTP, objects seemingly can be used to reason even about the most opaque code.}

\section{Object-Oriented Theorem Proving}

To introduce OOTP, it should be first and foremost noted that OOTP generalizes
and subsumes TTP. In particular, OOTP generalizes the notions of axioms,
theorems and inference rules of TTP to \textbf{simultaneous axioms},
\textbf{simultaneous theorems} and \textbf{simultaneous inference rules},
where simultaneity of a defined set of axioms, theorems or inference rules
allows \emph{interdependence} between the simultaneously-defined axioms,
theorems or inference rules (i.e., they can be defined
\emph{mutually-recursively}). In the sequel we explain in more
detail how OOTP is defined as a generalization of TTP.

In TTP, each axiom is a primitive theorem, while each inference rule
is a function from theorems to theorems~\cite{Paulson1996}. In OOTP, we have each \emph{set} of simultaneous axioms as a set of simultaneous primitive
theorems, while each set of simultaneous inference rules is an \textbf{object}
that maps---i.e., whose methods map---simultaneous theorems to simultaneous theorems.

Further, in TTP (as first suggested by Milner, for LCF~\cite{LCFPlotkin77,ML-LCF78}) regular FP type checking ensures that theorems could
be made only by axioms and inference rules. Applying inference rules
to already known theorems constructs proofs, rule by rule, in the
forward direction~\cite{Paulson1996}.  In OOTP, on the other hand, regular OO type checking
will ensure that simultaneous theorems could be made only by simultaneous
axioms and simultaneous inference rules. Applying simultaneous inference
rules to already known simultaneous theorems constructs simultaneous
proofs (as \emph{one} unit, called an \textbf{OO proof}, or an \textbf{object}), rule
by rule, in the forward direction.

Moreover, tactics in TTP permit a more natural style, backward proofs.
A tactic in TTP is a \emph{function} from goals to subgoals, justified by
the existence of an inference rule going the other way. The tactic
actually returns this inference rule (as a function) in its result:
tactics are higher-order functions~\cite{Paulson1996}. Similarly, in OOTP we have \textbf{OO
tactics}, which permit a more natural style, backward OO proof. An OO tactic
is an \emph{object} that maps---i.e., whose simultaneously-defined methods
map---(simultaneous) goals to (simultaneous) subgoals, justified by
the existence of a set of simultaneous inference rules going the other
way (i.e., from subgoals to goals). The OO tactic actually returns
this set of simultaneous inference rules (as an object) in its result:
OO tactics are higher-order objects.

Finally, tacticals in TTP provide control structures for combining
simple tactics into complex ones. The resulting tactics could be combined
to form still more complex tactics, which in a single step could perform
hundreds of primitive inferences. Tacticals are even more higher-order
than tactics~\cite{Paulson1996}. Similarly, \textbf{OO tacticals} in OOTP provide control
structures for combining simple OO tactics into complex ones. The
resulting OO tactics could be combined to form still more complex
OO tactics, which in a single step could perform hundreds of primitive
OO inferences. OO tacticals are even more higher-order than OO tactics.

Table~\vref{tab:OOTPvsTTP} summarizes the similarities and differences between OOTP
and TTP.

\begin{table}
\noindent \begin{centering}
\begin{tabular}{|c|c|}
\hline 
TTP & OOTP\tabularnewline
\hline 
\hline 
\multirow{2}{*}{Axiom} & Simultaneous Axioms\tabularnewline
&  (OO Axiom) \tabularnewline
\hline 
\multirow{2}{*}{Theorem} & Simultaneous Theorems\tabularnewline
& (OO Theorem)\tabularnewline
\hline 
\multirow{2}{*}{Inference Rule} & Simultaneous Inference Rules\tabularnewline

&  (OO Inference Rule)\tabularnewline
\hline 
\multirow{2}{*}{Proof (Function)} & Simultaneous Proofs\tabularnewline
&  (OO Proof; Object) \tabularnewline
\hline
Tactic & OO Tactic\tabularnewline
(Higher-order function) & (Higher-order object)\tabularnewline
\hline 
Tactical & OO Tactical\tabularnewline
(Higher-order function) & (Higher-order object)\tabularnewline
\hline 
\end{tabular}
\par\end{centering}

\protect\caption{\label{tab:OOTPvsTTP}Main notions in TTP and corresponding
notions in OOTP.}
\end{table}

\section{Remaining Work}
After overcoming some initial unfamiliarity it is easy to recognize
that, when formalized, OOTP, as presented
so far, seems to be a straightforward, relatively-mild generalization
of TTP. It should be noted, however, that object-oriented programming
(from which OOTP gets inspiration) is characterized by two features
that make it distinct and intricate, and thus---we
believe---more practically useful than functional programming (from which
TTP gets inspiration). These two features are (1) \emph{encapsulation} and
(2) (\emph{type}) \emph{inheritance} and \emph{subtyping}.

\subsection{Self-awareness and Inheritance}
Encapsulation, in OOP, first and foremost means the binding together of a collection
of related methods and fields in one unit (a capsule/an object). In addition, 
encapsulation in OOP refers to the
autognosticism property of objects (i.e., their self-awareness, via the
parameter \code{\emph{this}}, or \code{\emph{self}}, that is passed
implicitly to all methods of an object). Encapsulation in OOP also
frequently refers to the
ability to hide some implementation details. (The information hiding sense
of encapsulation is not characteristic to OOP, however, since it is supported in
FP, for example, via abstract data types.)

Type inheritance, and subtyping, in OOP refer to the inheritance and reuse of object types
and of \emph{behavioral contracts} by subclasses (and their corresponding subtypes)
from superclasses (and their corresponding supertypes), and to the preservation and
refinement of types and contracts in the subclasses/subtypes.

Providing a mathematical and practically-useful meaning of
encapsulation in OOTP (which we partially did above) and of inheritance and subtyping in OOTP (which yet remains to be done, but which we expect to involve some notion of \emph{specification inheritance} and/or \emph{theorem refinement}) should make OOTP a more intricate and more powerful generalization of TTP.

\subsection{Object-Oriented Software Verification (OOSV)}
We predict that a precise and full definition of encapsulation and inheritance in OOTP will immediately enable establishing an obvious and direct method for reasoning about object-oriented software and verifying its properties. Specifying the details of this method is work that remains to be done.

\section{Comparisons and Some Philosophy}

\subsection{OOTP versus TTP ... and OOSV}

Just as OOP is more practically useful than FP (as demonstrated by
the practical popularity of OO programming languages as compared to
that of functional programming languages\footnote{The skeptic reader should check by themselves the TIOBE index and langpop.com, or do an online survey to compare the popularity of OOP with that of FP. We also invite the reader to check our related work~\cite{AbdelGawad2018}.}), we believe OOTP might also prove to be
more practically useful in proving theorems about the real-world---intrinsically
complex, interconnected and interdependent?---than TTP is. In particular, as hinted to above
we believe OOTP, given its inherent object-orientation, can be used to reason about and verify OO software (OOSV) in a more natural manner.

\subsection{OOP versus FP}
In this section we briefly discuss the relation between OOP, as a computer programming paradigm centered around objects, and FP, as a paradigm centered around functions, by presenting an analogy and some philosophy. We offer what we believe is a novel perspective, but we do not go into many details given that the topic has been discussed, extensively but inconclusively, in the research literature (e.g., as revealed by~\cite{MacQueenMLOO02} and related work, or by an online search for `the expression problem').
\paragraph{An Analogy.}
Based on our modeling of nominal OOP and structural OOP~\cite{NOOPsumm}, in our opinion the main and most fundamental difference between OOP and FP is that OOP embraces mutual recursion wholeheartedly (mutually recursive methods come essentially for free in OOP)  while FP embraces mutual recursion reluctantly (mutually recursive functions require special constructs in FP, and even then their use is discouraged\footnote{See the code example in Appendix~\ref{app:Imp-OO} for an FP warning \emph{against} some definitions of mutually-recursive \emph{non-imperative} functions.}).

To an extent this difference is analogous to the
difference between natural languages that consider \emph{words} as
the main units of the language\footnote{Word-based languages; most natural
languages fall in this category since a word---a finite sequence of
letters---is the smallest unit that has meaning in such
languages.} and natural languages that consider \emph{characters} (or letters)
as the main fundamental units of speech and language\footnote{Character-based
languages; very few natural languages fall in this category in which
most if not all single characters, or ``letters,'' have meanings.}.

We find this analogy appropriate because in OOP an object, at its very
core, is a set of functions (the methods of the object, possibly \emph{mutually recursively}
defined). That is, in OOP an object is a finite set of tightly related functions
but each method in an object, on its own, is \emph{not} considered as
a fundamental independent unit (i.e., in OOP rarely, or at least less-frequently,
does a single method have an independent meaning, outside the context of
an object) but rather the grouping and binding of the set of functions (as methods)
in one unit (as an object, via the notion of \code{this}/\code{self}) is the fundamental, smallest-meaningful unit
in OOP. Functional languages, on the other hand, consider it to be the
norm that each function has an independent meaning, and they allow 
mutually-recursive functions usually only as a rare (and ``complex'') exception.

As such, akin to the view of words and letters in most
natural languages, OOP considers
mutually-dependent functions as the norm (and thus embraces their grouping as
a fundamental unit: an objects) while considering single functions having
independent meanings as being an exception (i.e., only very few objects have single meaningful methods in an OOP program, e.g., the \code{equals()} method of class \code{Object}\footnote{Even \code{equals()} is not quite independent. Its definition usually requires to be ``in sync'' with the definition of \code{hashCode()}~\cite{bloch08}.}). Functional programming,
on the other hand, similar only to very few natural languages, considers functions
having independent meanings as the norm (as we explained above).

\paragraph{A Philosophical Note.}
Consider the following three questions:
\begin{enumerate}

\item Is FP superior to OOP (as some PL researchers and theoreticians may claim), or vice versa
(as some industrial software developers may claim)?
\item To which of the two programming paradigms belongs the future of computer programming?
\item Is our world fundamentally composed of interdependent and complex entities (OOP),
or is it fundamentally composed of independent and simple entities (FP)?
\end{enumerate}

We believe these three questions
to be somewhat philosophical questions, whose answers will always remain controversial, rather than being technical questions whose answers can be
resolved scientifically. While having our own personal answers
to these questions (squarely in favor of OOP), we hope the brief discussion
above---including the illustrating analogy with natural languages---sheds
some light on why we believe these questions to be more philosophical and controversial questions and lesser technical and scientific ones.\footnote{We personally believe research on FP should aim to suggest how OOP can be improved---as was done in Java 8.0, for example---but that it should \emph{not} aim for FP eventually replacing OOP as the most widely used computer programming paradigm.}$^,$\footnote{Our opinion is formed based on nothing related to any software
efficiency arguments, in spite of their importance. We believe efficiency
arguments are fundamentally ambivalent, agnostic or neutral towards the `OOP
versus FP' question. Efficiency of computer software, particularly time
and space efficiency, is almost always due to the existence of imperative
programming features in a programming language. Imperative programming
features, which are usually found in OOP but not FP languages, are
not inherent or characteristic of OOP, and can be added to a FP language
with little additional complexity to the language (as ML---a popular
functional language---demonstrates).}

\section{Ideas to Incorporate Later}
\begin{itemize}
%\item Mutually recursive functions/SOOP.
%\item Convert code in~\cite[p.58]{Paulson1996} to an example object.
\item Use category theory~\cite{AbdelGawad2017b} to express and present OOTP?
\item Structured programming motivated Velleman's style of mathematical theorem proving that he calls `structured proving'~\cite{Velleman06}. Similarly, object-oriented programming motivated us to define OOTP.
\item In his widely read book `Road to Reality'~\cite[p.378]{Penrose2004}, Penrose writes
\begin{quote}
	It is my own personal opinion that we shall find that computability issues will eventually be found to have a deep relevance to future physical theory, but only a very little use of these ideas has so far been made in mathematical physics.
\end{quote}
Due to our belief in the strong interconnectedness of the real-world and given that not all mutual recursion (also called indirect recursion) can be translated to direct recursion~\cite{Kaser1993,Yu1997}, in affirmation to Penrose's opinion we believe that mutual recursion is one of the computability ideas that may be a key idea in physical theory, and, accordingly, that more use of mutual recursion (and of mutually recursive proofs, i.e., object-oriented proofs) in mathematical physics needs to be made.\footnote{As well as \emph{mutual corecursion}. Check our upcoming article on \emph{mutual coinduction}~\cite{AbdelGawad2019c}.}
\end{itemize}

\section{Implementation/Proof of Concept}

We are currently working on further exploration and refinement of
our ideas on OOTP by implementing a basic OO theorem prover in three
programming languages: ML (FP), Java (OOP) and Scala (OOP/FP), using
theorem prover ideas from our implementation of Paulson's Hal~\cite{Paulson1996}
and our update~\cite{PD2} to Velleman's Proof Designer~\cite{Velleman06,Vellem2006}.
Early versions of our code should be available on sourceforge
and github in the near future. Stay tuned.

\bibliographystyle{plain}

\newpage
\appendix
\section{Opaque Code Translated to OOP and FP Code}
\label{app:Imp-OO}
\subsection{Imperative to OOP}
The imperative code
\begin{lyxcode}
\begin{lstlisting}[language=C,tabsize=4]
var x := 0; y := 0; z := 0;
F:  x := x+1; goto G
G:  if y<z then goto F else (y := x+y; goto H)
H:  if z>0 then (z := z-x; goto F) else stop
\end{lstlisting}
\end{lyxcode}
can be translated to the following non-imperative OO code
\begin{lyxcode}
\begin{lstlisting}[language=Java,tabsize=4]
class C {
  final x, y, z: int
	
  // constructor
  C(xx,yy,zz: int) { x = xx; y = yy; z = zz }

  C F() { new C(x+1,y,z).G() }

  C G() { 
    if y < z then this.F()
    else new C(x,x+y,z).H() }

  C H() {
    if z > 0 then new C(x, y, z-x).F()
    else this }
}
\end{lstlisting}
\end{lyxcode}
Calling \code{new C(0,0,0).F()} returns an object equivalent to \code{new C(1,1,0)}, as expected.\\
\\
The code can be also translated to the following (more succinct, but harder to reason about) \emph{imperative} OO code
\begin{lyxcode}
\begin{lstlisting}[language=Java,tabsize=4]
class IC {
  x, y, z: int
		
  IC(xx,yy,zz: int) { x := xx; y := yy; z := zz }

  void F() { x := x+1; G() }

  void G() {
    if y < z then F()
	else { y := x+y; H() }}

  void H() { if z > 0 then { z := z-x; F() }}
}
\end{lstlisting}
\end{lyxcode}
On an object \code{ic = new IC(0,0,0)} calling \code{ic.F()} terminates with \code{ic} equivalent to \code{new IC(1,1,0)}, as expected.

It should be noted that in the OO code above all unqualified field references and methods calls are \emph{implicitly} done via \code{this}.\\

\subsection{Imperative to FP}

For comparison purposes, the corresponding mutually-recursive functional translation (adapted from the code in~\cite[p.58]{Paulson1996}) of the opaque imperative code above looks like
\begin{lyxcode}
\begin{lstlisting}[language=ML,tabsize=4]
type int_triple = int * int * int

fun F(x,y,z: int): int_triple =
      G(x+1,y,z)  
and 
    G(x,y,z: int): int_triple =
      if y<z then F(x,y,z)
      else H(x,x+y,z)
and 
    H(x,y,z: int): int_triple =
      if z>0 then F(x,y,z-x)
	  else (x,y,z);
\end{lstlisting}
\end{lyxcode}
Calling \code{F(0,0,0)} returns \code{(1,1,0)}.\\
\\
The functional code makes us note that
\begin{itemize}
\item All variables are passed explicitly as parameters in the code. With many variables, the code may not scale well.

\item The code, and functional programming more generally, do \emph{not} support \emph{open recursion}. Method calls in OOP are \emph{virtual} (i.e., are made via \code{this} at runtime, which is sometimes called dynamic/late-binding) by default in most OO languages. Function calls in FP, however, are always non-virtual (i.e., are always statically-bound). Unlike in OOP, FP lacks a notion of inheritance/subclassing that allows for late/dynamic-rebinding. An alternative to open recursion in FP is to make use of higher-order functions---i.e., functionals---that take other functions, dynamically, as parameters).

\item Finally, and most importantly, because referential transparency is frequently
stated as a main advantage of functional programming, Paulson~\cite[p.58]{Paulson1996}, in an implicit warning \emph{against} mutually-recursive functional programming (after presenting code similar to the functional code above, notably \emph{non-imperative}), remarks that
\begin{quote}
Functional programs are referentially transparent, yet can be totally opaque. If your code starts to look like this, beware!
\end{quote}
It should be noted that mutual recursion, and even open mutual recursion, is allowed and embraced wholeheartedly in OOP, due to the seemingly natural and intuitive real-world modeling capabilities of mutual recursion.\footnote{Consider, for example, the following list, from Webster's English dictionary, of words that are synonymous or near-synonymous to `mutuality': interdependence, interchange, reciprocity, symbiosis, give-and-take, cross-fertilization, and dialogue. Words in the list are strongly suggestive of (lively) interaction (between methods of an object)---a psychological and social factor that we conjecture could help explain the human appeal---and popularity---of OOP in comparison to FP.} It is a main goal of OOTP to make reasoning about such code (i.e., non-imperative but mutually-recursive code, expressed as OO code) as natural and straightforward (``transparent'') as TTP makes it for (non-imperative and non-mutually-recursive) functional code.
\end{itemize}
\end{document}